\documentclass[runningheads]{llncs}
\usepackage{graphicx}

\usepackage{amsmath}
\usepackage{amssymb}
\usepackage{bbding}
\usepackage{pifont}
\usepackage{wasysym}
\usepackage{utfsym}
\usepackage{fontawesome}
\usepackage{url}
\usepackage{subcaption}
\usepackage{mathtools}
\usepackage{color}
\usepackage{xcolor}
\usepackage[colorlinks=true,linkcolor=blue,citecolor=blue,urlcolor=blue]{hyperref}
\usepackage[misc]{ifsym}

\begin{document}
\title{Spatial-Division Augmented Occupancy Field for Bone Shape Reconstruction from Biplanar X-Rays}

\titlerunning{SdAOF}
%

\author{
Jixiang Chen\inst{1}\orcidID{0000-0001-9941-8324} \and 
Yiqun Lin\inst{1}\orcidID{0000-0002-7697-0842} \and 
Haoran Sun \inst{3}\orcidID{0009-0007-7390-2733} \and
Xiaomeng Li\inst{1,2(\textrm{\Letter})}\orcidID{0000-0003-1105-8083}
}
\authorrunning{J. Chen et al.}

\institute{
The Hong Kong University of Science and Technology \\
\email{eexmli@ust.hk} \and
HKUST Shenzhen-Hong Kong Collaborative Innovation Research \\ Institute, Futian, Shenzhen \and
Koln 3D Technology (Medical) Limited Company, Science Park, Hong Kong
}
%

%
\maketitle              
\begin{abstract}
Retrieving 3D bone anatomy from biplanar X-ray images is crucial since it can significantly reduce radiation exposure compared to traditional CT-based methods.
Although various deep learning models have been proposed to address this complex task, they suffer from two limitations: 1) They employ voxel representation for bone shape and exploit 3D convolutional layers to capture anatomy prior, which are memory-intensive and limit the reconstruction resolution. 2) They overlook the prevalent occlusion effect within X-ray images and directly extract features using a simple loss, which struggles to fully exploit complex X-ray information.
To tackle these concerns, we present Spatial-division Augmented Occupancy Field~(SdAOF). SdAOF adopts the continuous occupancy field for shape representation, reformulating the reconstruction problem as a per-point occupancy value prediction task. Its implicit and continuous nature enables memory-efficient training and fine-scale surface reconstruction at different resolutions during the inference.
Moreover, we propose a novel spatial-division augmented distillation strategy to provide feature-level guidance for capturing the occlusion relationship. 
Extensive experiments on the pelvis reconstruction dataset show that SdAOF outperforms state-of-the-art methods and reconstructs fine-scale bone surfaces.
The code is available at {\tt\small \url{https://github.com/xmed-lab/SdAOF}}.

\keywords{Bone Reconstruction \and Biplanar X-Rays \and Occupancy Field \and Implicit Neural Representation}
\end{abstract}

\newcommand{\yq}[1]{{\color{red}[#1]}}

\section{Introduction}

3D reconstruction of patient-specific bone anatomy is a crucial task for various clinical applications, such as biomechanical load analysis, surgical planning, and customized implant design optimization~\cite{paxton2022capturing,yamazaki2021patient,dreischarf2016estimation}. The current gold standard reconstruction of bones primarily involves segmentation from computed tomography (CT) images. However, CT acquisition involves relatively high ionizing radiation and is often unaffordable for many health centers~\cite{shakya2024benchmarking}. In contrast, X-ray images are more accessible due to lower costs and reduced scanning time.  Therefore, there is a significant need for the direct reconstruction of bone anatomy from a limited number of X-ray images. In this paper, we study the problem of bone mesh reconstruction from biplanar X-ray images.

Bone reconstruction from biplanar X-ray images tackles the challenging task of extracting 3D information from only two projection views, typically along the anterior-posterior (AP) and lateral view directions. Effectively modeling anatomical information is crucial for solving this ill-posed problem. Traditional approaches use statistical shape models~(SSM) or statistical shape and intensity models~(SSIM) to encode shape prior, but they suffer from cumbersome pre-modeling steps and sensitivity to model initialization~\cite{reyneke2018review,ehlke2013fast,zhang20133,clogenson2015statistical}. With recent advances in deep learning techniques, several learning-based approaches are proposed~\cite{almeida2021three,bayat2020inferring,buttongkum20233d,chen2020using,chenes2021revisiting,shakya2024benchmarking,chen2023bx2s,kasten2020end}. These methods primarily utilize 3D convolutional layers to generate volumetric shape model, while adopting varied fusion strategy to address the dimensionality misalignment between 2D input and 3D output.
However, the memory-intensive nature of voxel representation limits the reconstruction resolution and hinders the extraction of fine-scale bone meshes. An alternative approach for direct mesh reconstruction is to employ continuous implicit functions to represent the surface. This representation showcases superior efficiency for both model training and inference in different applications~\cite{saito2019pifu,lin2023learning,lin2024c,lin2024learning}. Inspired by this, we introduce the Occupancy Fields~(OF) representation~\cite{he2020geo}. This function indicates whether a 3D point lies inside the surface and defines the surface as an iso-surface of the continuous function. Based on this representation, we reformulate the reconstruction problem as the occupancy value prediction task for any 3D point within the target reconstruction space. This field can be captured using an OF Network~\cite{saito2019pifu}, which leverages pixel-aligned features from view-specific 2D feature planes and predicts per-point value through multi-layer perceptrons (MLP). This formulation facilitates fine-scale reconstruction in two respects: 1) The implicit representation does not require explicit storage, which is memory-efficient. 2) Its continuous nature allows us to scale into different resolutions during inference without retraining the model.
\raggedbottom

Despite leveraging the OF representation facilitates fine-scale mesh reconstruction, directly extracting features from input X-rays struggles to fully exploit information within complex X-ray images. The inherent complexity arises from the overlapped information in X-ray images, primarily due to common occlusion scenarios where the target bone is obscured by other tissues or its frontal parts. This occlusion stems from sequential attenuation along the view direction during the X-ray imaging process. While the occluded region of bones may vary due to different patient conditions, the anterior-posterior arrangement of organs follows a prior pattern when viewed from a fixed perspective. Therefore, we argue that it is possible to identify the occlusion pattern and further enhance the reconstruction quality. To achieve this, we propose a spatial-division augmented distillation strategy. It divides the target reconstruction space along each view direction into several subspaces and uses the Digitally Reconstructed Radiography (DRR) technique to render an augmented X-ray image for each subspace. A teacher OF network is trained to extract subspace-specific features and capture occlusion relationships across subspaces from these X-ray images. These features are then employed to provide feature-level guidance, enabling the identification of occlusion patterns in the original input X-rays through knowledge distillation.

To this end, we present Spatial-division Augmented Occupancy Field~(SdAOF), which integrates continuous implicit function representation and spatial-division augmented distillation strategy.  In contrast to voxel-representation-based methods, SdAOF demonstrates memory efficiency during training and is scalable to different resolutions during inference, achieving state-of-the-art fine-scale bone shape reconstruction results.
In summary, our main contributions include
\begin{enumerate}
    \item To the best of our knowledge, we are the first to introduce the continuous occupancy field representation for biplanar bone shape reconstruction. 
    \item We argue that identifying the occlusion relationship within X-ray images is crucial for robust feature extraction. To this end, we propose a spatial-division augmented distillation strategy to capture this information.
    \item We propose a novel framework SdAOF, which reconstructs fine-scale bone surfaces and is scalable to different resolutions without retraining.  
\end{enumerate}
\section{Methodology}

\subsection{Occupancy Field Network}
We use the continuous occupancy field~\cite{saito2019pifu,mescheder2019occupancy} representation, where the occupancy value $o(p)$ of a 3D point $p \in \mathbb{R}^3$ indicates whether it lies inside the surface (1) or not (0). The complete set of continuous occupancy values across all points constitutes the OF. We can generate the mesh surface by densely sampling the field over the space and extracting the iso-surface of $o(p)=0.5$ using the Marching Cube algorithm~\cite{lorensen1998marching}.
Therefore, the surface reconstruction is reformulated as per-point occupancy value prediction.

To this end, we introduce the OF network for bone reconstruction from multi-view X-rays, consisting of a feature extractor $\mathcal{F}$ and multi-layer perceptrons (MLP) for occupancy value prediction. Specifically, $\mathcal{F}$ extracts view-specific 2D features $\{ F_i \}_{i=1}^M \subset \mathbb{R}^{C\times H \times W}$ for $M$ input views~(particularly $M=2$ for biplanar reconstruction). Each 2D feature plane is regarded to be located at the X-ray detector plane of each respective view. For a 3D point $p$, we can obtain its pixel-aligned feature by first finding the projected location on each plane, and then querying the feature via interpolation. This is formulated as
\begin{equation}
\label{eq:query}
    f_i(p) = \text{Interp}(F_i, \pi(p)), 
\end{equation}
where $f_i(p) \in \mathbb{R}^C$ denotes the point-specific feature of the $i$-th view, $\text{Interp}(\cdot)$ denotes the bilinear interpolation operation, and $\pi(\cdot)$ refers to the projection operation. Multi-view per-point features are then concatenated and sent into MLP for occupancy value prediction, as formulated by 
\begin{equation}
\label{eq:predict}
    \hat{o}(p) = \text{MLP}\left(\text{Cat}(f_1(p), \ldots, f_M(p))\right). 
\end{equation}

The OF network uses mean-square-error (MSE) to calculate the per-point difference as the reconstruction loss
\begin{align}
    \mathcal{L}_{\text{recon}} = \frac{1}{N}\sum_{j=1}^N\left(\hat{o}(p_j) - o(p_j)\right)^2,
\end{align}
where $N$ denotes the number of points used for training and $o(p_i)$ denotes the ground-truth occupancy value.

\subsection{Spatial-division Augmentation}
Identifying the view-specific occlusion pattern helps to extract robust features related to bone anatomy. Although the occluded region of bones may vary due to different patient conditions, the anterior-posterior arrangement of organs follows a prior pattern when viewed from a fixed perspective. Therefore, we propose spatial-division augmentation to separate this arrangement information along view-depth dimensions and render augmented X-ray images for each input view. 

\begin{figure}[t]
    \centering
    \includegraphics[width=\linewidth]{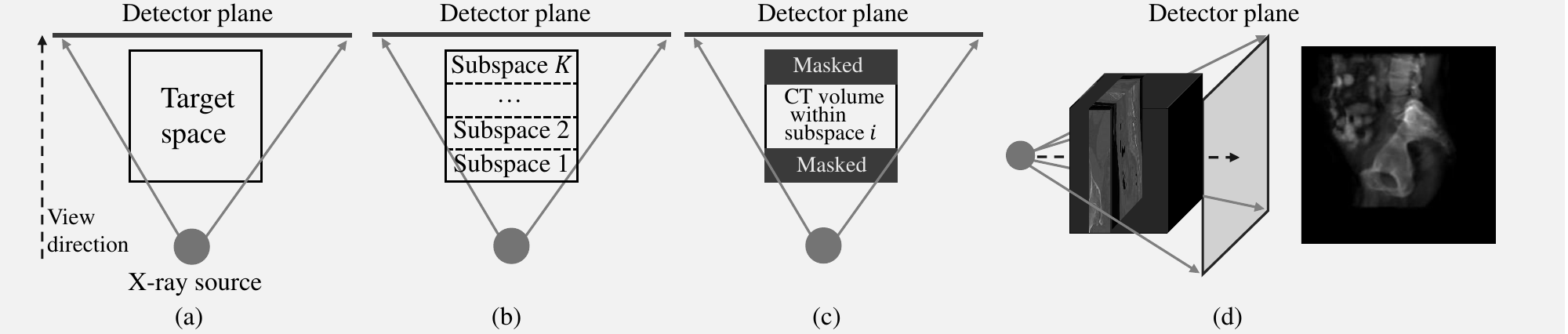}
    \caption{Illustration of spatial-division augmentation for a single view. (a)-(c): Planform of X-ray cone-beam scanning. (b): The target reconstruction space is divided along the view-depth direction into distinct subspaces. (c)-(d): For each subspace, we mask the regions of the original CT volume that lie outside the subspace and use DRR to generate an augmented X-ray image.}
    \label{fig:spatial-division}
\end{figure}

As depicted in Fig~\ref{fig:spatial-division}, we divide the target 3D space along the view direction into a set of $K$ subspaces for each view. Subsequently, within each subspace, we utilize the corresponding CT image by masking information outside of it and use DRR to generate $K$ augmented X-ray images, denoted by $\{ X^{\text{aug}}_{k}\}_{k=1}^{K}$. These X-ray images serve as the inputs of our teacher OF network. 

\subsection{Teacher Occupancy Field Network}
Since the OF network processes each input view identically except for the MLP prediction, in the subsequent illustrations, we explain the process for a single view input and omit the view index for brevity.
For each view, the teacher OF network extracts $K$ 2D feature maps $\{F^{\text{aug}}_k\}_{k=1}^{K}$ from $\{ X^{\text{aug}}_{k}\}_{k=1}^{K}$, where each feature map solely reflects information within the corresponding subspace. Following this, the view-specific feature of $p$ consists of $K$ folds $\{f_k^{\text{aug}}(p)\}_{k=1}^K$. To capture the anterior-posterior arrangement along the view direction, we fuse these features based on the view-depth relationship. For this purpose, we consider each feature plane to be located in the middle of its subspace and perform a weighted sum to aggregate features, as formulated below:
\begin{equation}
    f^{\text{aug}}(p) = \sum_{k=1}^{K} w_k \cdot f^{\text{aug}}_{k}(p),~\text{and}~
    w_k = \frac{1 / | d - d_k |}{\sum_{j=1}^{K} (1 / | d - d_j |) },
\label{Eq:depth-weighted-fusion}
\end{equation}
where $d_k$ denotes the view-depth distance of the $k$-th feature plane and $d$ refers to the view-depth distance of the projected point $p$. Intuitively, Eqn.~(\ref{Eq:depth-weighted-fusion}) assigns a larger weight to features that are closer to the point along the view-depth direction. Therefore, a point should identify its occlusion relationship based on feature planes closer to itself rather than the ones further away. The multi-view $f^\text{aug}(p)$ is then used as the input of MLP for reconstruction using $\mathcal{L}_{\text{Recon}}$.

\begin{figure}[t]
    \centering
    \includegraphics[width=\linewidth]{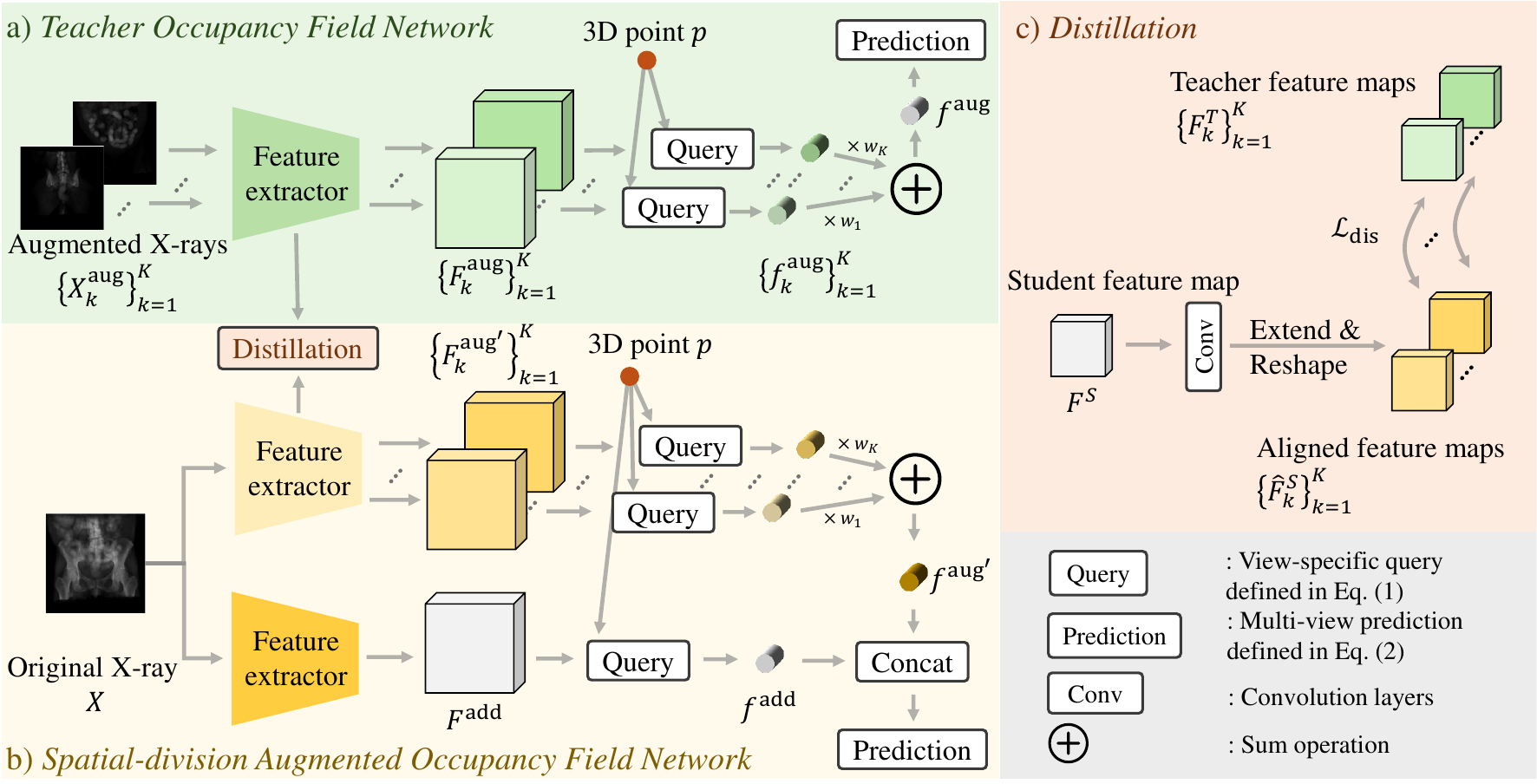}
    \caption{Overall framework of SdAOF. Both input views undergo the same processing, and we demonstrate the procedure using one input view as an example.}
    \label{fig:framework}
\end{figure}

\subsection{Spatial-division Augmented Occupancy Field Network}
SdAOF aims to identify the occlusion patterns under the teacher OF network's guidance from original X-ray inputs. As depicted in Fig.~\ref{fig:framework}, SdAOF employs two feature extractors, referred to as the student extractor $\mathcal{F}^S(\cdot)$ and the additional extractor $\mathcal{F}^\text{add}(\cdot)$. They each extract a 2D feature map from the original X-ray images $X$, denoted by $F^{S}$ and $F^{\text{add}}$, respectively. For each view, $\mathcal{F}^S(\cdot)$ performs feature-level distillation from the teacher OF network. The subspace-specific teacher features $\{ F_k^T\}_{k=1}^K$ consist of $K$ feature maps, which are extracted from $\{ X_k^{\text{aug}} \}_{k=1}^K$. In contrast, the student feature map $F^S$ solely includes one fold. To facilitate subspace-aligned knowledge transfer, we employ additional convolution layers to increase the channel of $F^S$ by $K$ times~($(c,w,h) \rightarrow (Kc,w,h)$), following by reshaping it for alignment~($(Kc,w,h) \rightarrow (K, c,w,h) $). SdAOF then performs feature-level distillation as follows
\begin{equation}
    \mathcal{L}_{\text{dis}} = \frac{1}{K}\sum_{k=1}^K \Big\| F_k^T - \hat{F}_k^S\Big\|^2,
\end{equation}
where $\hat{F}^S$ refers to $K$ extended student features and $\|\cdot\|$ indicates $L_2$ Norm for high-dimensional tensors. These features undergo the same fusion strategy as defined in Eqn.~(\ref{Eq:depth-weighted-fusion}), generating $f^{\text{aug}'}(p)$.
On the other hand, the extracted $F^{\text{add}}$ explores additional information that complements the occlusion-aware ones. Its corresponding point feature $f^{\text{add}}(p)$ is obtained following Eqn.~(\ref{eq:query}).
SdAOF then concatenate $f^{\text{aug}'}(p)$ and $f^{\text{add}}(p)$ as the view-specific point feature for MLP inputs. The overall loss for training SdAOF includes the distillation objective and the reconstruction objective
\begin{align}
    \mathcal{L} = \mathcal{L}_{\text{Recon}} + \alpha \cdot \frac{1}{M L}\sum_{i=1}^M\sum_{j=1}^{L} \mathcal{L}_{\text{dis}},
    \label{overall_loss}
\end{align}
where $\alpha$ balances the contribution of two loss terms. In practice, the distillation is performed at $L$ intermediate layers. During inference, only the SdAOF network is used, and the teacher OF network is disregarded.
\section{Experiments}
\subsection{Experimental Settings}

\noindent \textbf{Dataset and Preprocessing. }
We conduct extensive experiments on the Pelvis dataset from the XrayTo3Dbenchmark~\cite{shakya2024benchmarking} to validate the effectiveness of our proposed method in reconstructing bone shapes from biplanar X-rays. The original segmentation annotations are utilized to generate fine-scale meshes, serving as ground truth. We specifically exclude some noisy fragment annotations and omit the sacrum part due to insufficient quality in its annotation for fine-scale mesh derivation. The data split follows the original setting, with 321 samples used for training, 57 for validation, and 67 for testing. 2D X-ray images are generated by DRRs using the TIGRE~\cite{biguri2016tigre} package under the cone-beam geometry. Please refer to supplementary material for implementation details.

\begin{table}[t] 
    \centering
    \caption{Comparison results on XrayTo3Dbenchmark-Pelvis dataset. Mean and standard deviation results are reported. $\ast$: Evaluate at $336^3$ output resolution.}
    \label{tab:sota}
    \resizebox{\textwidth}{!}{
    \begin{tabular}{c|c|c|c|c|c}
    \hline
        Method & \multicolumn{2}{c|}{Left pelvis} & \multicolumn{2}{c|}{Right pelvis} & \\
    \hline
     & CD (mm) $\downarrow$ & EMD (mm) $\downarrow$ & CD (mm) $\downarrow$ & EMD (mm) $\downarrow$ & Inference time (s) \\
    \hline
    \multicolumn{6}{c}{Voxel-based methods}  \\
    \hline
    UNet~\cite{kasten2020end} & $2.939 \pm 0.778$ & $7.203 \pm 4.373$ & $2.995 \pm 0.623$ & $6.779 \pm 3.652$ & $0.004$ \\
    UNetr~\cite{hatamizadeh2022unetr} & $3.194 \pm 0.716$ & $8.651 \pm 4.309$ & $3.294 \pm 0.684$ & $8.934 \pm 5.015$ & $0.015$ \\
    SwinUNetr~\cite{hatamizadeh2021swin} & $2.817 \pm 0.578$ & $6.181 \pm 2.076$ & $2.887 \pm 0.462$ & $6.355 \pm 2.212$ & $0.018$ \\
    3DReconNet~\cite{buttongkum20233d} & $2.628 \pm 0.433$ & $5.817 \pm 1.856$ & $2.675 \pm 0.369$ & $6.085 \pm 1.810$ & $0.016$ \\
    AttentionUNet~\cite{oktay2018attention} & $2.842 \pm 0.562$ & $6.426 \pm 2.205$ & $2.949 \pm 0.708$ & $6.579 \pm 2.631$ & $0.005$ \\
    \hline
     \multicolumn{6}{c}{Implicit-representation-based methods (\textbf{Ours})} \\
    \hline
    PIFu~\cite{saito2019pifu} & $2.586 \pm 0.418$ & $5.899 \pm 1.623$ & $2.623 \pm 0.291$ & $\textbf{5.943} \pm \textbf{1.529}$ & $0.033$ \\ 
    SdAOF & $\textbf{2.465} \pm \textbf{0.367}$ & $\textbf{5.710} \pm \textbf{1.818}$ & $\textbf{2.571} \pm \textbf{0.290}$ & $6.160 \pm 1.526$ & $0.033$ \\
    \hline
    $\text{PIFu}^{\ast}$~\cite{saito2019pifu}  & $2.494 \pm 0.404$ & $4.985 \pm 1.601$ & $2.407 \pm 0.335$ & $\textbf{4.502} \pm \textbf{1.385}$ & $8.822$ \\ 
    $\text{SdAOF}^{\ast}$ & $\textbf{2.335} \pm \textbf{0.359}$ & $\textbf{4.556} \pm \textbf{1.692}$ & $\textbf{2.321} \pm \textbf{0.329}$ & $4.612 \pm 1.480$ & $11.852$ \\
    \hline
    \end{tabular}
    }
\end{table}

\noindent \textbf{Baseline Methods. }
We select 6 publicly available methods for comparison, including 5 state-of-the-art voxel-based methods: UNet~\cite{kasten2020end}, UNetr~\cite{hatamizadeh2022unetr}, SwinUNetr~\cite{hatamizadeh2021swin}, AttentionUNet~\cite{oktay2018attention}, 3DReconNet~\cite{buttongkum20233d} and one representative implicit-representation-based method in human-shape reconstruction, PIFu~\cite{saito2019pifu}. As PIFu has not been applied to this task previously, we implemented it ourselves using the original code base. We follow the configuration in ~\cite{shakya2024benchmarking} to use an input resolution of $128 \times 128$ and evaluate with a resolution of $128^3$. For voxel-based methods, we derive the occupancy voxels and use the Marching Cube algorithm~\cite{lorensen1998marching} to extract the mesh surface.

\noindent \textbf{Evaluation Metrics. }
We follow prior works~\cite{saito2019pifu} to use the Chamfer Distance (CD) and Earth Mover's Distance (EMD) to assess reconstructed surface accuracy. Each method undergoes three runs to calculate the metrics. Lower CD/EMD values represent better higher reconstruction quality. Detailed metric explanations are provided in the supplementary material.

\begin{figure}[t]
    \centering
    \includegraphics[width=\linewidth]{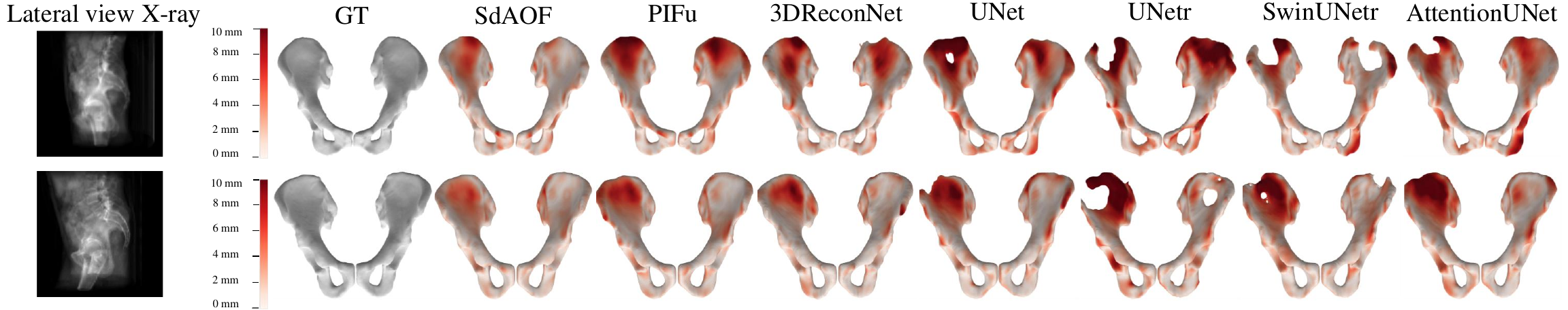}
    \caption{Quantitative comparison results evaluated at $128^3$ output resolution. The color indicates the distance between the reconstructed mesh and the GT mesh. More quantitative results are provided in supplementary material. The corresponding lateral view X-rays show the occlusion pattern.}
    \label{fig:quantitative}
\end{figure}

\subsection{Results}
\begin{figure}[tb]
  \begin{minipage}[t]{0.48\textwidth}
    \centering
    \captionof{table}{Ablation on spatial-division augmented strategy.}
     \resizebox{\linewidth}{!}{
    \begin{tabular}{c|c|c|c|c}
    \hline
    \multicolumn{3}{c|}{Settings} & \multicolumn{2}{c}{CD (mm)} \\
    \hline
    Index & Spatial division & Distillation & Left pelvis & Right pelvis \\
    \hline
    A) & \ding{55} & \ding{55} & 2.586 & 2.623 \\ 
    B) & \ding{51} & \ding{55} & 2.703 & 2.699 \\
    C) & \ding{55} & \ding{51} & 2.547 & 2.612 \\
    D) & \ding{51} & \ding{51} & \textbf{2.465} & \textbf{2.571} \\
    \hline
    \end{tabular}
    }
    \label{tab:ablation}
  \end{minipage}%
  \hfill  
  \begin{minipage}[t]{0.48\textwidth}
    \centering
    \caption{GPU memory usage comparison when training with a batch size of 1.}
    \includegraphics[width=0.8\linewidth]{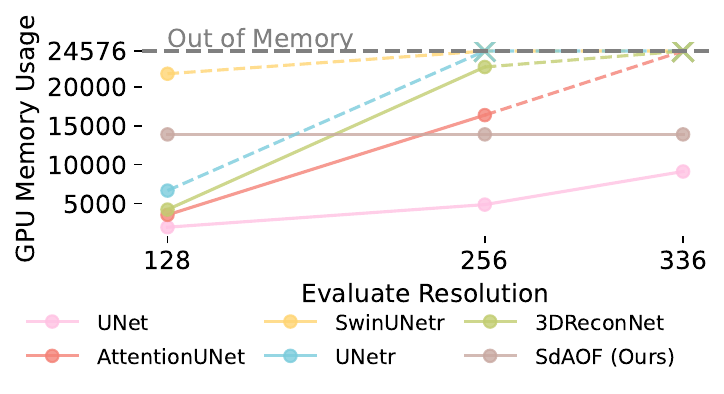}
    \label{fig:memory}
  \end{minipage}
\end{figure}

\noindent
\textbf{Comparison with State-of-The-Arts.}
As indicated in Table~\ref{tab:sota}, our method outperforms voxel-based approaches, achieving a substantial reduction in reconstruction error by $6\%$ and $4\%$ compared to the previously leading method (3DReconNet) for the left and right pelvis, respectively. The left pelvis, positioned distantly from the viewing source (from right to left in the lateral view), faces challenges of occlusion by the right pelvis. SdAOF leverages feature-level guidance to identify occlusion patterns, resulting in a notable improvement in left pelvis reconstruction, as illustrated in Fig.~\ref{fig:quantitative}. Furthermore, methods based on implicit representation offer scalability to different output resolutions, facilitating fine-scale reconstruction. The results evaluated at $336^3$ output resolution demonstrate SdAOF's further enhancement in reconstruction accuracy ($2.465 \rightarrow 2.335$ mm of CD on the left pelvis) while maintaining a reasonable reconstruction time (11 seconds per sample). Additionally, the training memory comparison results in Fig.~\ref{fig:memory} underscore the superiority of SdAOF. Unlike voxel-based methods that commonly struggle to train models at fine-scale resolutions, SdAOF's resolution scalability avoids the memory bottleneck associated with explicit representation.

\noindent
\textbf{Ablation Studies.}
The ablative analysis results of spatial division information and the distillation strategy are presented in Table~\ref{tab:ablation}, with the following settings:
A): The vanilla PIFu baseline;
B): Instead of guiding the model to capture spatial division information via distillation, we train an image-to-image translation network~\cite{isola2017image} that generates augmented X-ray images from original X-ray images. These predicted images are then used as additional inputs of SdAOF. The performance degrades as the generation network brings more noise than additional subspace-related information in the generated augmented X-rays.
C): We train the teacher model by only inputting original X-ray images that are identical to the student's inputs. Then, we perform distillation using these features. The result highlights the crucial role of spatial-division information in improving reconstruction performance. Without such information, distillation degenerates into a vanilla ensemble of encoders, and the improvement is marginal.
D): Our proposed SdAOF achieves the best result, validating the effectiveness of both the spatial division information and the distillation strategy.

\begin{figure}[t]
    \centering
    \includegraphics[width=0.8\linewidth]{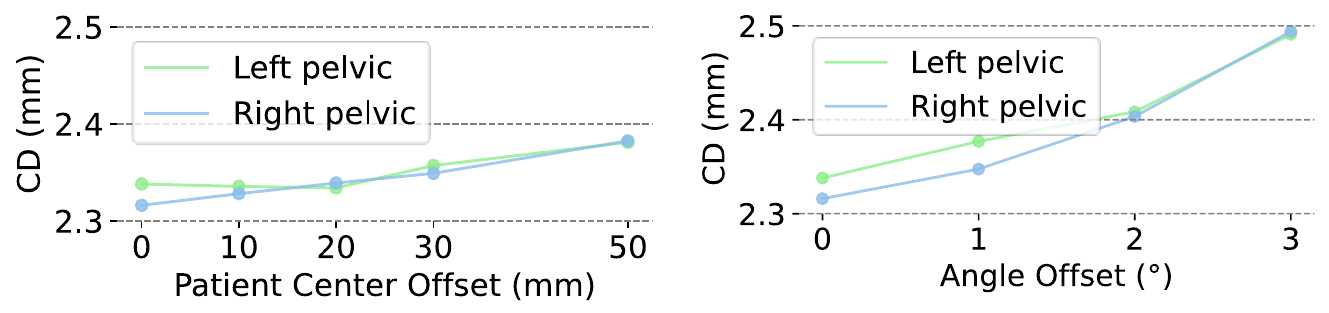}
    \caption{Results on different imaging parameters at $336^3$ output resolution.}
    \label{fig:robustness}
\end{figure}

\noindent 
\textbf{Robustness to different imaging parameters.}
In real-world scenarios, the imaging direction may deviate from the perfect alignment with AP and lateral views, and the patient's center may vary. These deviations cause the magnification effect within X-ray images to differ from the training distributions. The results in Fig. \ref{fig:robustness} show that SdAOF is robust to minor variations in view angles and patient center locations.

\section{Conclusion}
In this study, we introduce SdAOF, a novel framework designed for the challenging task of reconstructing bone shapes from biplanar X-ray images. SdAOF achieves state-of-the-art fine-scale reconstruction results by integrating the continuous occupancy field representation and the spatial-division augmented distillation strategy. The proposed SdAOF is applicable to multi-view scenarios (whether single view or more than two views), offering adaptability across various settings to balance scanning cost, radiation exposure, and reconstruction quality. Our findings highlight the reconstruction benefits obtained from capturing occlusion patterns. Future research avenues could explore more generalized methods, such as incorporating adaptive subspace division based on 3D semantics. This promising direction remains a focus for our future work.

\subsubsection{\ackname}
This work was supported in part by grants from the National Natural Science Foundation of China under Grant No. 62306254, grants from the Hong Kong Innovation and Technology Fund under Projects PRP/041/22FX and Project of Hetao Shenzhen-Hong Kong Science and Technology Innovation Cooperation Zone (HZQB-KCZYB-2020083).

\subsubsection{Disclosure of Interests.}
The authors have no competing interests to declare that are relevant to the content of this article.
\newpage
\bibliographystyle{splncs04}
\bibliography{miccai_bib}
\end{document}


\noindent \textbf{Spatial-Division Augmented Occupancy Field for Bone Shape Reconstruction from Bi-Planar X-Rays -- Supplementary Material}
\setcounter{figure}{4}
\setcounter{equation}{6}
\setcounter{table}{2}
\begin{figure}[h]
    \centering
    \includegraphics[width=0.9\linewidth]{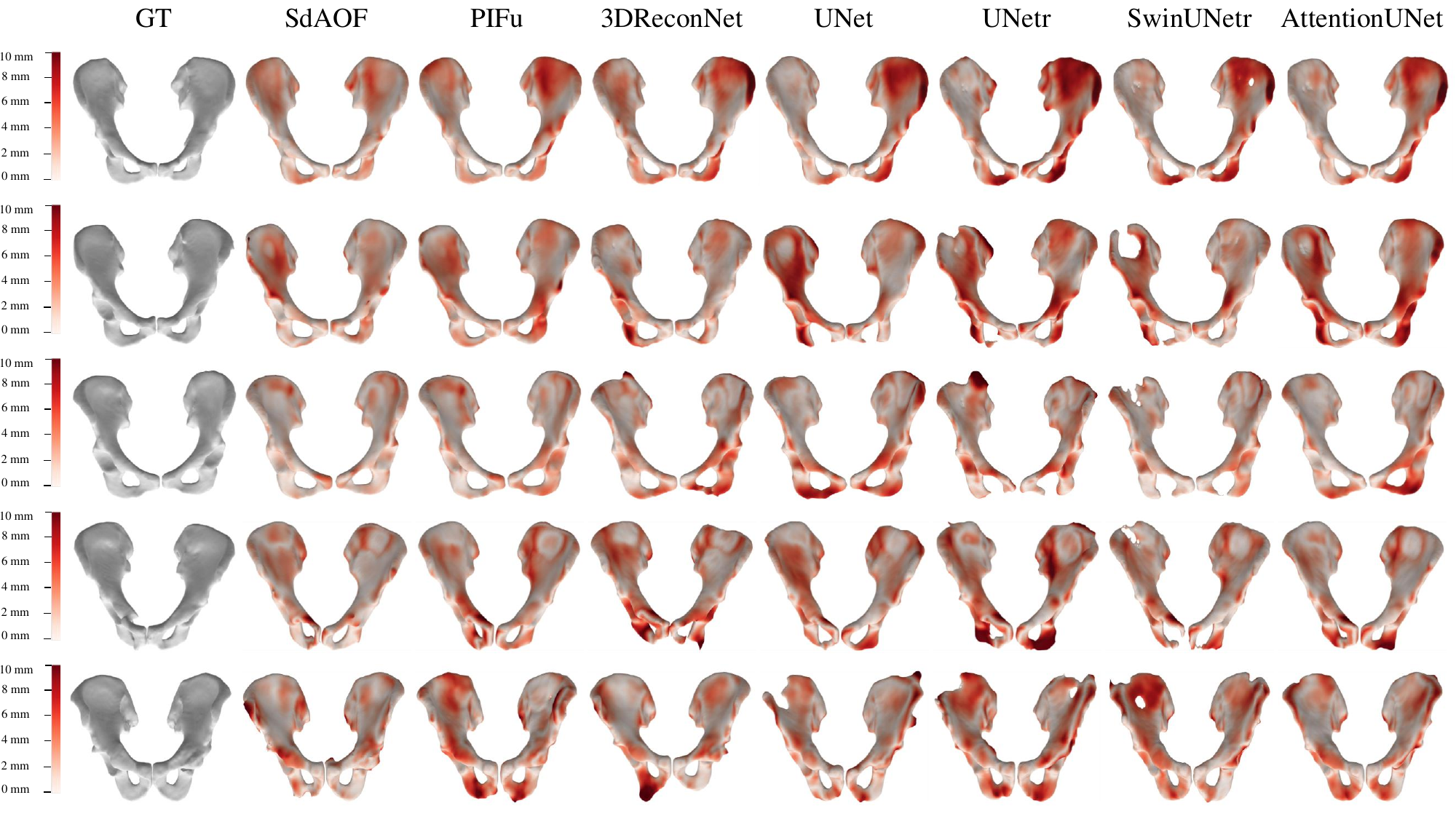}
    \vspace{-0.3cm}
    \caption{More quantitative comparison results evaluated at $128^3$ output resolution.}
    \label{fig:enter-label}
\end{figure}


\section{Implementation Details}
\noindent \textbf{Voxel-based Methods.} We implement the 5 voxel-based representation using the code implementation in ~\cite{shakya2024benchmarking}. They have already configured proper parameters for each method via hyperparameter search, including the structure (channel, depth) settings, learning rate, batch size, and optimizer.

\noindent \textbf{Implicit-representation-based Methods.} We implement SdAOF using PyTorch with a single NVIDIA RTX 3090 GPU. Here we introduce key implementation details. Please refer to supplementary material for more configurations. We use UNet as the feature extractor with output feature channel $C=128$. The output features of each double convolution layer, excluding the input convolution, contribute to distillation, resulting in $L=9$ layers utilized for this purpose.  The subspace division number is set to $K=4$. During each forward pass, $N=10000$ points are sampled. Half of these points are uniformly distributed across the space, while the remainder is sampled from a region within 2mm of the surfaces. Given that each occupancy field can represent only one watertight bone surface, we set the output channel to 2 for deriving the left and right pelvic surfaces. The loss balance term $\alpha$ in Eqn.~(\ref{overall_loss}) is empirically set to 0.2. Both PIFu and SdAOF use the same set of hyperparameters. Specifically, we use the Adam optimizer with an initial learning rate of 5e-3. The learning rate is linearly warmed up from 5e-5 to 5e-3 during the first 25 epochs, then decays by multiplying a factor of 0.8 after at $\{50, 100\}$ epoch. Both PiFU, teacher OF Network, and SdAOF are trained for 110 epochs. The MLP for occupancy value prediction consists of 6 layers with skip connection (channels:  $512 \rightarrow 1024 \rightarrow 512 \rightarrow 256 \rightarrow 128 \rightarrow 2$).

\section{Evaluation Metrics}
The Chamfer Distance~(CD) and Earth Mover's Distance~(EMD) measures dissimilarity between two point sets. In mesh reconstruction context, $N$ points are uniformly sampled from the predicted mesh surface and ground-truth mesh surface for calculation, denoted by $S_{\text{pred}}, S_{\text{GT}} \subseteq \mathbb{R}^3$. In practice, we set $N=4096$.

\noindent \textbf{Chamfer Distance.}
CD measures both the reconstruction accuracy and completeness by calculating the average distance from each point in one set to its nearest neighbor in another set and vice versa. It is defined as 
\begin{equation}
    \text { CD }(S_{\text {pred }}, S_{\text{GT}}) = \frac{1}{2} ( \frac{1}{\left|S_{\text {pred }}\right|} \sum_{p \in S_{\text{pred}}} \min _{q \in S_{\text{GT}}}\|p-q\|_2 + \frac{1}{\left|S_{\text {GT }}\right|} \sum_{q \in S_{\text{GT}}} \min _{p \in S_{\text{pred}}}\|p-q\|_2).
\end{equation}

\noindent \textbf{Earth Mover's Distance.}  Earth Mover's Distance (EMD) measures a better perceptual similarity assessment compared to other distance measures. Given two sets, one can be seen as a mass of earth properly spread in space, the other as a collection of holes in that same space. EMD calculates the minimum work required to fill the holes with earth. The work unit corresponds to a unit of Euclidean distance. Its calculation is formulated by
\begin{equation}
    \text{EMD}(S_{\text{pred}}, S_{\text{GT}}) = \frac{1}{\left| S_{\text{pred}} \right|} \min_{\phi : S_{\text{pred}} \rightarrow S_{\text{GT}}} \sum_{p \in S_{\text{pred}}} \| p - \phi(p) \|_2,
\end{equation}
where $\phi : S_{\text{pred}} \rightarrow S_{\text{GT}}$ is a bijection.

\begin{figure}[t]
    \centering
    \includegraphics[width=0.8\linewidth]{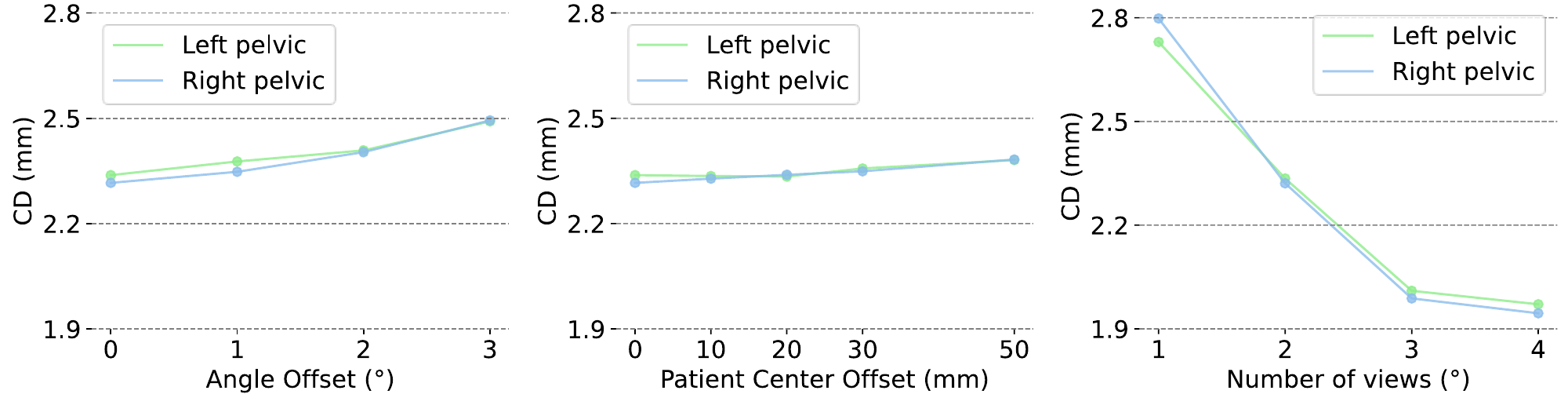}
    \vspace{-0.2cm}
    \caption{Additional experiment results evaluate at $336^3$ output resolution.}
    \label{fig:additional_exp}
\end{figure}

\section{Additional experiments}
\noindent \textbf{Robustness to different imaging parameters.} In real-world scenarios, the imaging direction may deviate from the perfect alignment with AP and lateral views, and the patient's center may vary. These deviations cause the magnification effect within X-ray images to differ from the training distributions, which assume ideal bi-planar view directions and iso-center patients. The results presented in Fig. \ref{fig:additional_exp} demonstrate that SdAOF exhibits robustness to minor variations in both view angles and patient center locations.

\noindent \textbf{Extension to multi-view input.} The proposed SdAOF can naturally extend to multiple view directions. The right part of Fig. \ref{fig:additional_exp} illustrates the CD of SdAOF when using 1 to 4 input views (with the AP view as the single input view). These results highlight that SdAOF consistently enhances the reconstruction quality with an increasing number of input views. This flexibility provides a trade-off between radiation dose and improved accuracy in various application scenarios.
\bibliographystyle{splncs04}
\bibliography{miccai_bib}